\documentclass[preprint]{aastex}
\usepackage{amsmath}

\def\Msun{\, M_\odot}
\def\Lsun{\, L_\odot}

\def\kms{\, {\rm km\, s^{-1}}}
\def\pc{\, {\rm pc}}
\def\cm{\, {\rm cm}}
\def\au{\, {\rm AU}}
\def\yr{\, {\rm yr}}
\def\sgra{{\rm Sgr ~ A*}}

\slugcomment{ApJ, Accepted June 26 2012}
\shorttitle{Disrupted Star in Galactic Center}
\shortauthors{Miralda-Escud\'e}

\begin{document}

\title{
A star disrupted by a stellar black hole as the origin
of the cloud falling toward the Galactic center}
\author{
Jordi Miralda-Escud\'e\altaffilmark{1,2}
}
\altaffiltext{1}{Instituci\'o Catalana de Recerca i Estudis Avan\c cats,
                 Barcelona, Catalonia
}
\altaffiltext{2}{Institut de Ci\`encies del Cosmos,
                 Universitat de Barcelona/IEEC, 
                 Barcelona, Catalonia
}

\begin{abstract}
We propose that the cloud moving on a highly eccentric orbit near the central
black hole in our Galaxy, reported by Gillessen et al., is formed by a
photoevaporation wind originating in a disk around a star that is tidally
perturbed and shocked at every peribothron passage. The disk is proposed to
have formed when a stellar black hole flew by the star, tidally disrupted its
envelope, and placed the star on its present orbit with some of the tidal
debris forming a disk. A disrupting encounter at the location of the observed
cloud is most likely to be caused by a stellar black hole because of the
expected dynamical mass segregation; the rate of these disk-forming encounters
may be as high as $\sim 10^{-6}$ per year. The star should also be spun up by
the encounter, so the disk may subsequently expand by absorbing angular
momentum from the star. Once the disk expands up to the tidal truncation radius,
the tidal perturbation of the outer disk edge at every peribothron may place
gas streams on larger orbits which can give rise to a photoevaporation wind that
forms the cloud at every orbit. This model predicts that, after the cloud is
disrupted at the next peribothron passage in 2013, a smaller unresolved cloud
will gradually grow around the star on the same present orbit. An increased
infrared luminosity from the disk may also be detectable when the peribothron
is reached. We also note that this model revives the encounter theory for planet
formation.
\end{abstract}

\section{Introduction}

  Drawing on the vast technological advance in adaptive optics, infrared
detectors and X-ray telescopes, observations over the last two decades
have revolutionized our knowledge of the Galactic center region. It is
now beyond reasonable doubt that a black hole of mass $4\times 10^6
\Msun$ is present at the center of the Milky Way surrounded by a stellar
cusp with a total mass in stars of $\sim 10^6 \Msun$ within the central
parsec \citep[for a review, see][]{G10}. Several massive young
stars are present in the central 0.1 pc with well determined orbits, many
of which are part of a disk structure and were born in a starburst
$\sim$ 6 Myr ago. Hot gas is also present throughout this region,
believed to originate from the stellar winds of these massive stars.
%producing a luminosity of $\sim 0.5 \Lsun$ in
%X-rays and $\sim 100 \Lsun$ in the submillimeter in the central 0.02 pc.

  A mysterious cloud of gas was also reported recently by \citet{G12},
moving along a Keplerian orbit. The cloud emits $5\Lsun$ of
continuum infrared light interpreted as dust emission at a temperature of
550 K, and hydrogen and helium recombination lines consistent with a
photoionized cloud with a gas temperature of $10^4$ K. The cloud is
being rapidly disrupted as shown by a clear velocity gradient along its
resolved long axis of $\sim$ 100 AU, which is consistent with the tide
along its highly eccentric orbit: the cloud has fallen from an
apobothron at $\sim$ 8000 AU and will pass through peribothron at $\sim$
250 AU in summer 2013.

  The origin of this cloud is a most intriguing question. Three
possibilities have been proposed so far. In the first one, the cloud is
isolated and diffuse and was formed by the collision of stellar winds
from massive stars \citep{B12,S12}. Stars near the inner edge of the
disk at a distance $r=8000$ AU from the black hole, where the circular
velocity is $v_c\simeq 700 \kms$, may emit winds at velocities near
$v_c$ which can collide and leave material at low velocity. This
material might cool after being shocked at high density and form a
shell, and then fall on the observed highly eccentric orbit. This model
faces some difficulties: clouds produced from wind collisions should
have a very high velocity dispersion (resulting from the wind and the
star velocities added in quadrature), so it appears unlikely that only
one prominent cloud is observed which needs to have formed with a very
low velocity at the inner edge of the disk of young stars. There are no
clear candidates among the known young stars with winds that might have
produced the required cloud near the observed apocenter. The cloud needs
to cool down and be confined by the pressure of the external hot medium
because its self-gravity is negligible, and it is not clear how the
cloud may have avoided fragmentation through the Kelvin-Helmholtz and
Rayleigh-Taylor instabilities as it moved through the hot medium from
its apocenter. In addition, the presence of dust in a cloud at $10^4$ K
that has cooled from gas shocked to millions of degrees after being
ejected in a wind from a hot star is also difficult to account for.

  A second possibility might be that an evolved star is losing mass that
is producing the cloud at every orbit, which might be observed as a
planetary nebula were it not tidally removed at every peribothron
passage. The star would need to be very hot and not highly luminous, and
the dust abundance should be very low, to avoid reradiating too much of
the stellar light in the infrared and be consistent with the upper limit
on the $K$-band flux ($K$-band absolute magnitude above 1) and the observed
flux at longer wavelengths \citep{G12}. This model also has
severe problems: in order to be hot enough and faint enough, the star
should be smaller than a solar radius, and any wind would then be too
fast to produce the observed cloud over the orbital period of 130 years.
Furthermore, with no more than $\sim 10^4$ low-mass stars expected
within 0.04 pc from the Galactic center, the chances of catching this
one at a very brief and rare stage of its stellar evolution, while no
other luminous old giant in a stage of longer duration has been
found as close to the center, must be very small.

  A third possibility is that the cloud is formed by photoevaporation of
a circumstellar disk around a star embedded in the cloud \citep{ML12}.
The circumstellar disk may have accreted from gas as usual when the star
was formed in the recent starburst. As long as the star is not very
massive its $K$-band flux is far below the observational upper limit
\citep{G12}. At an initial orbital radius of 8000 AU for the star, the
disk would be tidally limited to a size of $\sim$ 10 AU. \citet{ML12}
proposed that this disk gave rise to the observed cloud of 100 AU after
the star was deflected into the present orbit, leading to faster
photoevaporation and tidal stretching as the peribothron is approached.
The problem with this model is the difficulty in deflecting the star
from a low eccentricity orbit in the disk of young stars to the observed
highly eccentric orbit without disrupting the gaseous disk.

  This is easily seen by considering the example of an encounter with a
$m=10 \Msun$ star, typical among the objects dominating the dynamical
relaxation rate, moving with a relative velocity $\sigma \sim 200 \kms$
characteristic of the disk velocity dispersion, and at an impact
parameter $b=10$ AU. The velocity deflection caused in this encounter is
$\Delta v = 2Gm/(b\sigma) = 10\kms$, which is similar to the disk
circular velocity at 10 AU. Therefore the disk would lose a large
fraction of its mass for this impact parameter and would be destroyed in
closer encounters. However, deflecting the star from a disk orbit to the
observed highly eccentric orbit requires a velocity change $\Delta v\sim
500 \kms$, which is clearly impossible to achieve over a single orbit.
In fact, the probability for just one such encounter at $b\sim 10$ AU
over one orbit for any given star in the disk is much less than unity.

  Nevertheless, the possibility to produce the cloud from a
photoevaporating circumstellar disk around a low-mass star is an
interesting one, since it can naturally produce many of the observed
features of the cloud. It is therefore natural to ask if there are
other ways of producing a gas disk around a star in the environment
of the Galactic center.

  This paper proposes that a disk was formed when an old, low-mass star
suffered a close encounter with a stellar black hole, which tidally
disrupted its outer envelope and deflected the star into its present
orbit. Even though some of the tidal debris may have escaped the star, a
large fraction of the mass stayed bound and fell back to the star,
creating a small disk. The star was also spun up by the encounter, and
gradually transferred its angular momentum to the disk. The resulting
expanded disk can then create a cloud like the one observed at every
orbit. Most of the disk mass stays within the tidal radius of 1 AU at
the peribothron, while a small fraction migrates out to a larger radius
at every orbit, where it is photoionized and drives a wind that
generates the cloud.

  We shall first discuss the rate at which encounters of stars and
stellar black holes that can lead to substantial disruption and disk
formation should occur near the Galactic center in Section 2. The possibility
to create the observed cloud from the disk is considered in Section 3, and we
summarize the tests of the model and our conclusions in Section 4.

\section{Disrupting encounters of stars with stellar black holes}

\subsection{The density of stars and black holes}

  The most likely place in the Galaxy where a disrupting encounter
between a normal star and a stellar black hole may take place is near
the Galactic center, where both the density of stars and black holes are
highest. The density of stellar black holes should be particularly 
enhanced owing to migration by dynamical friction of the most massive
objects in the old stellar population of the bulge toward the center.
About $\sim 20000$ stellar black holes are estimated to have migrated to
the stellar cusp surrounding $\sgra$ over the age of the Galaxy
\citep{M93,MG00}. A stellar cusp undergoing dynamical relaxation with a
constant outflow of orbital energy should have a density profile
$\rho(r)\propto r^{-7/4}$ \citep{BW76,BW77}; the total profile may vary
when a range of stellar masses is present, but does not strongly deviate
from this form \citep[e.g.,][]{A09}. The population of old stars
dominates the contribution to the stellar mass at large radius, and
there is a critical radius $r_b$ within which the stellar black holes
dominate. Inside $r_b$, the density of stellar black holes, $\rho_b$,
probably approaches the 7/4 slope, and their profile becomes much
steeper outside $r_b$. The density of low-mass stars, $\rho_s$, probably
approaches a 3/2 slope inside $r_b$, corresponding to a constant phase
space density in the Keplerian potential, although the profile may be a
bit flatter if many stars are destroyed by collisions.

  Assuming that the total stellar mass roughly follows the 7/4 slope and
normalizing the profile to a total mass $10^6 \Msun$ within 1 pc
\citep{G10}, and if a total of 20000 stellar black holes with an average
mass of $M_b=10\Msun$ have migrated to this region, then a mass equal to
that of all the stellar black holes is contained within $r_b \simeq
0.3\pc$. Their number density at $r<r_b$ is
\begin{equation}
 n_b(r) = n_{b0}\, (r/r_b)^{-7/4} \qquad\qquad (r< r_b) ~.
\end{equation}
We use the normalization $n_{b0}=(5/16\pi) \, 5\times 10^3 r_b^{-3}$,
which assumes that 25\% of the stellar black holes are inside $r_b$,
while the other 75\% are outside following a steeper profile. The
remaining 75\% of the mass inside $r_b$ is treated here for simplicity
as a single population of main-sequence stars with mass $M_s=1\Msun$,
with a profile
\begin{equation}
 n_s(r)=36\, n_{b0}\, (r/r_b)^{-3/2} \qquad\qquad (r< r_b) ~.
\end{equation}
This simple model yields mass densities similar to those found in the
Fokker-Planck calculation of \cite{HA06}.

\subsection{Impact parameters for disk formation}

  In order to create a disk, an encounter needs to be close enough to
cause a strong tidal distortion and raise matter into orbit around the
star. While many numerical simulations of stellar collisions have been
carried out, starting with the work of \citet{BH87,BH92}, the question
of whether a disk can be formed from the tidal streams that fall back
toward the star after an encounter between two objects that remain
unbound has not received so much attention. The numerical simulations,
however, show a fraction of the tidally stripped mass forming a disk
structure immediately after the encounter (see, e.g., \citealt{L93} for
collisions of massive stars, and \citealt{K93} for a tidal interaction
with a black hole).

  The problem of disk formation after a tidal interaction is related to
the encounter theory for the formation of a planetary system, where
planets form after a plume of material is lifted from a star due to an
encounter with another star. It was pointed out by \cite{S39} that the
plume cannot condense directly into planets because of its high internal
pressure and long cooling time. However, the high internal pressure
together with the tidal forces from the perturbing object can provide a
lateral force that redistributes angular momentum in the plume, so that
some of the material that is left on bound orbits may form a disk
instead of falling back onto the stellar surface. The disk should
initially be very hot, but after the encounter it can in principle cool
over many orbits and eventually form planets. The problem is also
similar to the theory of formation of the Moon in a collision of two
planets, which has been studied in detail, and where it has been shown
that a disk can be formed containing a mass of more than $10^{-2}$ of
the mass of the two colliding planets \citep[see the review by][]{C04}.

  To estimate the required impact parameter for strong disruption in our
specific problem, we consider a star of mass $M_s$ and radius $r_s$
encountering a stellar black hole of mass $M_b$ with an initial relative
velocity $v_r$, at an impact parameter $b$ leading to a closest approach
at peribothron $r_p$. The velocity of the star at peribothron,
approximating the trajectory of its center of mass to be the same as for
a point particle, is $v_p = (v_r^2 + 2GM_b/r_p)^{1/2}$, and conservation
of angular momentum implies $r_p = b v_r/v_p$. We first consider the
case $v_r < v_0$, where we define
\begin{equation}
  v_0 \equiv v_e \left(M_b\over \sqrt{2}M_s \right)^{1\over 3} =
 \left[ G (2M_b)^{2/3} M_s^{1/3} \over r_s \right]^{1\over 2} ~,
\end{equation}
and $v_e=(2GM_s/r_s)^{1/2}$ is the escape velocity of the star. For this
case, we use the strong disruption condition that the tidal acceleration
caused by the black hole between the center and surface of the star
along the radial line at peribothron, at distances $r_p$ and $r_p-r_s$,
is equal to the gravitational acceleration on the surface due to the
star. This implies $2M_b r_s/r_p^3 = M_s/r_s^2$, or a maximum peribothron
distance for tidal disruption of
\begin{equation}
  r_p = r_s \left( 2M_b\over M_s \right)^{1\over 3} \qquad \qquad
  (v_r < v_0) ~.
\end{equation}
This condition agrees with the maximum impact parameter found in
numerical simulations required for stripping matter (e.g.,
\citealt{K93}; note that the approximation used in these simulations
that $r_p \gg r_s$ is only marginally correct for our case).
The condition $v_r < v_0$ ensures that the velocity at peribothron is
$v_p \simeq (2GM_b/r_p)^{1/2}=v_0$, and the duration of the strong tide
is $t\simeq r_p/v_p = \sqrt{2} r_s/v_e$, equal to the free-fall time of
the star. For the case $v_r > v_0$, gravitational focusing remains small
at $r_p$, and so $v_p\simeq v_r$ and the duration
of the strong tide is shorter than the star free-fall time by the factor
$v_r/v_0$. Our condition for strong distortion is in this case
$2M_b r_s/r_p^3 (r_p/v_r)= M_s/r_s^2 (\sqrt{2} r_s/v_e)$, or
\begin{equation}
  r_p = r_s \left( 2M_b\over M_s \right)^{1\over 3}\, \left( v_0\over v_r
 \right)^{1\over 2}  \qquad \qquad  (v_r > v_0) ~.
%  r_p = r_s \left( {v_e \over v_r}{\sqrt{2}M_b\over M_s} right)^{1/2}
% \qquad \qquad   (v_r > v_0) ~.
\end{equation}

  The corresponding maximum impact parameters are
\begin{equation}
  b = r_p {v_p\over v_r} \simeq r_s\left( 2M_b\over M_s \right)^{1\over 3} \,
 {v_0\over v_r}   \qquad \qquad  (v_r < v_0) ~,
\label{bm1}
\end{equation}
\begin{equation}
  b \simeq r_p \simeq r_s \left( 2M_b\over M_s \right)^{1\over 3}
 \left( v_0\over v_r \right)^{1\over 2}  \qquad (v_r > v_0) ~.
\label{bm2}
\end{equation}
We do not take into account a minimum impact parameter, even though the
star is totally disrupted when the impact parameter is sufficiently
small. The impact parameter required for a complete destruction should
be substantially smaller than the maximum values needed for raising
matter from the surface. The large degree of concentration of stars
implies that the dense core is hard to disrupt and may have a disk
forming around it even when a large fraction of the star is tidally
pulled out.

  For much larger velocities, $v_r > v_e (M_b/M_s)$, strong disruption
requires the black hole to cross through the star and becomes
inefficient as the tidal acceleration acts over a shorter time. Some
material may be dragged out of the star through the narrow cylinder that
the black hole perforates in these very fast encounters, but it would be
difficult for any disk to be formed.

\subsection{The rate of disrupting encounters}

  The rate of encounters at impact parameters smaller than the maximum
values for disruption in equations (\ref{bm1}) and (\ref{bm2}) can now
be calculated as
\begin{equation}
 R= 4\pi^2 \int dr\, r^2\, n_b(r)\, n_s(r)\, \left< b^2\, v_r \right> ~,
\label{rate1}
\end{equation}
where $v_r$ is the relative velocity between a star and a black hole,
and $\left< b^2\, v_r \right>$ is computed by averaging over the velocity
distributions at a given radius.

  Dynamical equilibrium implies that the rms one-dimensional velocity
dispersion of a set of particles moving in a Keplerian potential with a
density profile $n \propto r^{-\gamma}$ is $\sigma^2=GM/r/(\gamma+1)$.
Hence, the rms relative velocity of stars and stellar black holes at a
distance $r$ from the central black hole of the Milky Way of mass $M$
(referred to as $\sgra$) is
\begin{equation}
  \left< v_r^2 \right>\, = {3GM \over r} \left({2\over 5} + {4\over 11} \right)
  = {126 \over 55} {GM\over r} ~.
\end{equation}
There is a critical radius $r_0$ at which this rms relative velocity is
equal to $v_0$,
\begin{equation}
 r_0 = r_s {63\over 55} {M\over M_s} \left(
 { \sqrt{2} M_s\over M_b } \right)^{2\over 3} \simeq 6000\au ~.
\end{equation}
We also define the radius at which the rms relative velocity reaches
$v_e M_b/M_s$, within which disruptions become inefficient,
\begin{equation}
 r_f = r_s {63\over 55} {M\over M_s} \left( M_s \over M_b 
 \right)^2 \simeq 200\au ~.
\end{equation}
Approximating also $ \left< 1/v_r \right>\, \simeq
(\left< v_r^2 \right>)^{-1/2}$, the
total rate of encounters from equation (\ref{rate1}) is
\begin{equation}
  R = 4\pi^2 r_b^3\, 36 n_{b0}^2 r_s^2 v_e {\sqrt{2} M_b \over M_s} \,
 \times 
  \left[ \int_{r_f/r_b}^{r_0/r_b} dx\, x^{-{5\over 4}} \, +
  \left(r_0\over r_b \right)^{-{1\over 2}} \int_{r_0/r_b}^1 dx\,
 x^{-{3\over 4}} \right] ~.      
\end{equation}
The first integral arises from the outer region $r>r_0$, where slow
encounters affected by gravitational focusing dominate, and the second
integral is for $r< r_0$, where fast encounters limited by the duration
of the strongest tidal acceleration dominate. The result is,
\begin{equation}
  R = 576\pi^2 r_b^3 n_{b0}^2 r_s^2 v_e {\sqrt{2} M_b \over M_s} \,
 \times
 \left[ \left(r_b\over r_f\right)^{1\over 4}
      + \left(r_b\over r_0\right)^{1\over 2}
      - 2 \left(r_b\over r_0\right)^{1\over 4}
 \right] \simeq 10^{-6} \, {\rm yr}^{-1} ~.      
\label{rate2}
\end{equation}
The rate of interesting encounters in radial shells of constant
logarithmic width is fairly flat, but it is actually maximum at the
smallest radius, near $r_f$. If the observed cloud is indeed being
produced by a star that was tidally perturbed, it is interesting to note
that even though the encounter might have occurred at any point along
the present cloud orbit, the most likely place would be near the
peribothron, which is close to $r_f$. In this case, the black hole would
have rushed very close to the surface of the star at $\sim 6000 \kms$.

  The predicted rate of encounters implies that the perturbed star needs
to produce the observed cloud for many orbits in order to have a
reasonable probability to be observing the cloud at a random time.

\subsection{Effects of other types of collisions}

  A small disk around a star may also be formed as a result of a tidal
interaction or collision between two main-sequence stars. As for the case
of black holes, it is useful to divide these encounters into cases when
the relative velocity between the two stars is smaller or larger than
the escape velocity of the star. 

  Encounters with a relative velocity smaller than the escape velocity
take place mostly at large radius, and have effects that are dominated
by the tidal interaction. The impact parameters for which an important
amount of mass can be tidally raised from a star to form a disk are
therefore given approximately by equation (\ref{bm1}). The encounter
rates for a specific star is proportional to the number density of
perturbers times their mass, i.e. to the mass density of perturbers,
which is comparable for stars and stellar black holes at radii up to
$\sim r_b$. The total rates are therefore comparable, but encounters
among two stars are more likely to be produced at a radius much larger
than the orbit of the observed cloud.

  Encounters between two stars at radii smaller than $\sim 0.1$ pc
mostly occur at velocities higher than the escape velocity. In this
case, physical collisions are more important than tidal effects, and the
required impact parameters are about $2r_s$ independently of the
velocity. Stellar collisions may therefore become more frequent than
tidal encounters with stellar black holes at small radius, despite the
shallower density profile of stars. Most of the debris produced by the
direct physical collision in these high-velocity encounters would be
left on unbound orbits, but some of the mass may be pushed out of
the stars at low velocity, remain bound and also form a disk. Exactly
how much mass might be left on bound orbits can only be estimated with
detailed numerical simulations that are beyond the scope of this paper,
but in principle stellar collisions might substantially increase the
total rate of disk-forming encounters.

\section{Evolution of the stellar disk and wind}

  After a strongly distorting encounter, a large fraction of the mass of
the perturbed star may either be thrown out on unbound orbits, or may
eventually fall back to the star. The fraction of the stellar mass that
avoids these two outcomes and is left on bound orbits with enough
angular momentum to form a disk depends on many physical parameters, and
can only be obtained from detailed hydrodynamic simulations. Here, we
are going to assume as a characteristic value that the disk may have a
mass of $\sim$ 1\% of the stellar mass, a typical value in the case of
collisions among terrestrial planets that can account for the formation
of the Moon \citep{C04}.

  Immediately after the tidal encounter, the disk should be small
because most of the debris should not acquire a large specific angular
momentum. The star should be strongly spun up during the encounter
\citep[see][]{A01}, and afterward it should settle to an equilibrium
with an equatorial radius larger than its main-sequence value because of
fast rotation and the dissipation of energy into internal heat, which
will take a Kelvin-Helmholtz time ($\sim 10^7$ years) to be radiated
away.

\subsection{Required wind speed}

  In order to make the observed gas cloud, the star and disk need to
generate a wind with an adequate mass loss rate to deliver the mass of
the cloud over an orbital period, and at a velocity that is low enough
not to exceed the observed present size of $100 \au$. We simplify the
treatment of the motion of a gas element in the wind moving away from
the star by approximating the falling trajectory of the star from its
apobothron to its peribothron as if it were on a purely radial orbit
with zero orbital energy (the actual observed cloud is on an orbit with
eccentricity $e=0.94$). The distance $r$ from the star to $\sgra$ at
time $t$ is then
\begin{equation}
  r(t) = \left[ {3\over 2} \sqrt{2GM} (t_0-t) \right]^{2\over 3} ~,
\end{equation}
where $t_0$ is the time when the star would reach $r=0$ if it were in a
purely radial orbit. A gas element separating along the radial direction
at a distance from the star $x(t)\ll r$ is affected by a tidal
acceleration $g_t = 2GMx/r^3$. Neglecting the gravity of the star
(which is only important at an initial time when the wind is launched
from a small value of $x$) and any ram-pressure force due to the hot
medium around $\sgra$ (see Gillessen et al.\ 2012 and Burkert et al.\
2012 for a discussion of the effects of ram-pressure), the motion for
the gas element is described by the equation
\begin{equation}
  {d^2 x\over dt^2} = {2GMx \over r^3} = {4x\over 9 (t_0-t)^2} ~.
\end{equation}
Assuming that the gas element is at a distance equal to the observed
cloud size, $x_1\simeq 100 \au$, at the time $t_1\simeq t_0 - 2\yr$ of
the observations reported by \cite{G12}, and that it was emitted by the
wind from a distance $x \ll x_1$ near the time of the apobothron,
$t_a\simeq t_0 - 70 \yr$, the solution to the above equation is (using
$t_0 - t_a \gg t_0 - t_1$),
\begin{equation}
  x(t) = x_1 \left( t_0 - t_1 \over t_0 - t \right)^{1\over 3} 
   \left[ 1 - \left(t_0 - t \over t_0 - t_a \right)^{5\over 3} \right] ~.
\end{equation}
The initial velocity of the wind therefore should be about
\begin{equation}
  \dot x(t_a) = {5 x_1 (t_0-t_1)^{1/3} \over 3 (t_0-t_a)^{4/3} } \simeq
   4 \kms  ~.
\end{equation}

  Any wind that is generated from the small disk that is initially
formed after the tidal encounter of a star with a black hole would have
a velocity of hundreds of $\kms$ (not much smaller than the escape
velocity of the star), which is much too fast to explain the observed
cloud. To generate the required slow wind, a mechanism is needed to
expand the disk and to provide energy for launching a wind from large
radius.

\subsection{Disk expansion}

  The disk expansion may result from the fast rotation of the perturbed
star. Note that the star may already have been a fast rotator
before the encounter that created the disk, because previous encounters
with stellar black holes in the Galactic center region at larger impact
parameters (which occur more frequently) may have gradually spun up the
star \citep{A01}; the last, closest encounter may simply
have cracked up the rotation rate even further. After the encounter, a
process of angular momentum transfer from the star to the disk should
result in an expansion of the disk. If the star rotates very fast, it
may become prolate and cause a rotating gravitational tide on the inner
disk that can transfer the angular momentum. An oblate star that is
still rotating faster than the inner disk can continue to transfer
angular momentum if it is magnetically connected to the disk. The disk
will be spread by internal transport of angular momentum, pushing
matter on the outer edge to an increasing orbital size as more angular
momentum is acquired from the star on the inner edge.

  Let the angular momentum of the star after it has settled to
hydrostatic equilibrium following the encounter with the black hole be
$L_s = \phi_L \sqrt{GM_s^3 r_s}$. As an example, for a spherical object
with a singular isothermal density profile truncated at $r_s$, and a
surface rotation velocity equal to the circular orbital velocity,
$\phi_L=2/9$. The angular momentum is even larger for a prolate star
rotating near the maximum rate which has expanded owing to the increase
of internal energy (decrease in absolute value) in the tidal event.
The angular momentum of the disk of mass $M_d$ is $L_d=M_d
\sqrt{GM_s r_d}$, where $r_d$ is the characteristic disk radius
obtained from its mass-weighted average of $\sqrt{r}$. If the star
transfers most of its angular momentum to the disk, the final radius
of the disk is
\begin{equation}
 r_d = r_s \left(\phi_L M_s\over M_d \right)^2 ~.
\end{equation}
For $M_d= 0.1 \phi_L M_s$, the disk can expand out to $100 r_s$, or
$0.5 \au$ for a solar-type star.

  If, as proposed in this paper, this star and expanded disk system are
inside the observed cloud in the Galactic center, the disk cannot expand
beyond $r_d\sim 0.5\au$ because the tidal limit at the peribothron of the
cloud orbit, $r_{cp}$, is $r_t = r_{cp}\, [M_s/(2M)]^{1/3}\simeq 0.7
\au$, so the disk is truncated at this size at every orbital period of
140 years.

\subsection{Mass loss rate}

  The escape velocity from the surface of a disk at $r_d\sim 0.5\au$
is $\sim 60 \kms$, still too large to generate a slow wind at a
velocity of a few $\kms$. An additional mechanism is required to first
spread a small fraction of the gas in the disk over a larger region
around the star at every orbit, which can then be blown out at a low
velocity.  Moreover, photoionization from the massive stars near the
Galactic center can provide the energy required to generate the wind
once some material expands to the radius where the escape velocity is
reduced to near the isothermal sound speed at the temperature of
photoionized gas, $c_i\simeq 11\, (T/10^4\, {\rm K})^{1/2} \kms$ \citep{ML12}.
As the wind escapes the gravity of the star, its velocity
can be moderately reduced below $c_i$ to the value required to reproduce
the size of the observed cloud.

  The total mass loss rate from a cloud of radius $r_c$ that is being
photoevaporated by an external flux of ionizing photons $F_i$ can be
roughly estimated as $\dot M \sim 4\pi r_c^2 c_i n_e \mu_e$, where $n_e$
is the electron density in the external ionized layer that shields the
interior of the cloud, and $\mu_e$ is the mean mass per electron. The
condition that the ionizing flux is balanced by a recombination rate
column $\alpha_B n_e^2 \ell$, where $\alpha_B$ is the case B
recombination coefficient and $\ell \sim r_c$ is the length of the
ionized layer, is then used to estimate
$n_e \sim [F_i/(\alpha_B r_c)]^{1/2}$. A detailed calculation was
presented by Bertoldi \& McKee (1990), who obtained
\begin{equation}
 \dot M = 1.4\times 10^{-11} \phi_w \,{S_{49}^{1/2} \over d_{\rm pc}}\,
 r_{\au}^{3/2}\, \Msun\yr^{-1} ~,
\end{equation}
where $r_{\au} = r_c/(1\au)$ is the radius of the photoevaporating cloud
expressed in AU, $\phi_w$ is a dimensionless factor that is written as a
combination of other modeling dimensionless factors in equation (4.2) of
Bertoldi \& McKee (1990), and the external flux is expressed as $F_i =
10^{49}/(4\pi) S_{49}/d_{\rm pc}^2 \pc^{-2}$, with $S_{49}$ equal to the
total emission rate of ionizing photons in units of $10^{49}\,
{\rm s}^{-1}$ from a source at a distance $d_{\rm pc}$ expressed in parsecs.
Here, we assume that the stars in the young disk emit $S_{49}=10$ 
\citep[which is 15\% of all the ionizing luminosity in the central
0.5 pc; see][]{G10} from a typical distance of 0.06 pc, which yields
$S_{49}^{1/2}/d_{\rm pc} = 50$, or $F_i=2\times 10^{14}\cm^{-2}$. The
parameter $\phi_w$ depends on a photoevaporation parameter defined as
$\psi = \alpha_B F_i r_c/c_i^2$. Using $c_i=11 \kms$ (at $T=10^4$ K) and
$r_c= 5\au$, we find $\psi\simeq 3000$. The value of the dimensionless
factor is then $\phi_w\simeq 4$, as shown in Figure 11 of \cite{BM90},
and the inferred mass-loss rate is $\dot M\simeq 3\times
10^{-8} \Msun\yr^{-1}$ for a cloud size of $r_{\au} = 5$.

  Therefore, as long as a mass of at least $3\times 10^{-6}\Msun$
can be expelled from the disk after the star has passed by the
peribothron, and can reach out to a distance from the star $r_c\sim
5\au$, then this mass can be slowly lost from the system over an orbital
period of $\sim 100$ years, roughly at the desired wind speed to produce
the observed cloud. This amount of mass in a region of a radius
$r_c = 5 \au$ has a number density $n\sim 10^9 \cm^{-3}$, which is
self-shielded behind an ionized layer with $n_e\simeq 10^{6.5}\cm^{-3}$.

  How can this mass move from the disk out to $\sim 5 \au$ and stay
there for $\sim 100$ years? A possible way for this to happen is
discussed next.

\subsection{Generation of the photoevaporating cloud from the disk}

  As described previously, a mass of $\sim 10^{-2} M_s$ can reasonably
be placed in a disk and be transported outwards to a radius $\sim 0.5
\au$. A fraction of only $10^{-3.5}$ of this disk mass needs to be
ejected out to a large distance to generate a photoevaporation rate of
$3\times 10^{-8} \Msun \yr^{-1}$ over 100 years. If a mechanism to eject
this small fraction of the disk mass near the escape velocity exists,
this can in principle occur at every orbit
after the peribothron passage and create a similar cloud to the one
we observe for more than $10^3$ orbits, or a total time of more than
$10^5$ years. The rate of encounters between stars and black holes in
equation (13) would then imply a reasonably large probability of
observing one cloud at any random time near the Galactic center. This
requires the disk to expand and eject mass at every orbit in an optimal
way to produce the observed cloud, but even if the process is much less
optimal (with more mass being ejected and perhaps dispersed instead of
accumulating in the observed cloud), the probability to observe the
cloud at a random time may still be a reasonable one.

  We note that the inferred mass of the observed cloud is $M_c \simeq
10^{-5} f_V^{1/2} \Msun$, where $f_V$ is the filling factor of gas with
density $n_e\simeq 10^{5.5} f_V^{-1/2}$ in a spherical cloud with radius
$\sim 100 \au$ \citep{G12}. A filling factor $f_V\sim 0.1$ is
probably most reasonable, because the cloud is expected to have a
filamentary shape owing to the tidal acceleration that stretches the
cloud in the direction of the orbit and compresses it across both
perpendicular directions. Only the long axis of the cloud is
observationally resolved.

  The mechanism to eject a small fraction of the disk mass may occur
when the star-disk system reaches peribothron, and the disk undergoes
rapid precession and is strongly warped in its outer part by the tidal
forces. If the disk expands slowly as the star loses angular momentum, a
very small fraction of
the disk may diffuse outside the tidal radius during one orbit, but a
larger fraction may be present into the intermediate region near $0.5\au$
where the disk is not yet torn apart but is substantially warped and
perturbed, leading to collisions of gas streams and shocks that can
eject gas near the escape velocity. Inevitably, some of the ejected
mass will escape the system, but some may simply move out on a large
orbit and remain bound to the star. Material that is ejected near
the escape velocity from the disk just after the peribothron passage
can remain near a separation from the star where the tidal
acceleration from $\sgra$ is comparable to the gravitational attraction
of the star. Complex orbits are therefore possible that leave gas
streams far from the disk with enough angular momentum to prevent them
from falling back to the disk. Furthermore, lateral pressure forces
should also redistribute angular momentum in the gas moving away from
the disk, which is heated by the ambient ultraviolet light to
temperatures above $1000$ K even when hydrogen ionization is still
prevented by self-shielding. In a disk outflow that is non-spherical and
highly inhomogeneous, pressure gradients at this temperature can change
the velocity of gas streams by $\sim 3 \kms$, providing substantial
angular momentum.

  The scenario that this leads to is of a large region of turbulent gas
motions around the smaller disk, with random gas streams moving on
different orbits. Eventually these gas streams would collide and cool,
and if the net angular momentum of the gas is still small, most of the
gas should fall back to the disk. However, it may take several orbits
for this process to be completed, and gas streams at $\sim 5$ - 10
$\au$ from the star need only survive for a few orbits to produce a
steady wind that generates the cloud, until the next peribothron is
reached. In practice, a larger
amount of mass may come off the disk at every peribothron, but
the largest fraction of this may be launched on small orbits where it should
indeed cool and fall back to the central disk, while a
smaller amount of gas that moves out on larger orbits may suffice to
sustain the photoevaporation wind.

\section{Discussion}

  A model is proposed in this paper to explain the origin of the gas
cloud described by \cite{G12}. At some place along the present orbit of
the gas cloud, a close encounter of a star and a stellar black hole
occurred perhaps $10^4$ or $10^5$ years ago that strongly disrupted the
star, tearing out a substantial fraction of its mass into debris and
spinning up the star to near the break-up point. The fraction of the
debris that remained bound to the star either fell back on the star or
formed a small disk around it. The star, left on the orbit of the
present cloud, settled back to equilibrium as a fast rotator, perhaps
with a prolate shape initially. Subsequently, the disk gradually
expanded as it absorbed the angular momentum of the star, until its
outer edge reached the truncation radius at peribothron. After this
time, the disk has been launching a fraction of its mass at every
peribothron passage in gas streams that arise from the strong tidal
perturbation on the outer disk edge. Most of the streams remain on small
orbits and fall back to the disk shortly afterward, and other gas
becomes totally unbound from the star, but some of the gas streams on
intermediate orbits move out to 5 - 10 $\au$ of the star and form a
turbulent cloud. These turbulent streams complete only a few orbits
around the star before the next peribothron passage, and so they do not
have enough time to collide, cool and fall back to the disk. The
photoevaporation of these streams by the ambient ionizing radiation
generates a wind, which is elongated into a filament by the tidal force
as the star falls back to the peribothron on its next orbit and produces
a cloud like the observed one.

  The material that reaches the outer edge of the disk at $\sim 0.5
\au$ over the entire duration of the cloud-generating phenomenon may
exceed $10^{-2} \Msun$ for a rapidly rotating star. Several inefficiency
factors are likely to be present to convert this mass into the
photoionized clouds that are produced at every orbit: some mass may
rapidly escape the system after being ejected from the disk, and not all
the photoionized wind may follow the star in a single coherent cloud for
the whole orbit if hydrodynamic instabilities induced by ram-pressure
from the hot medium fragment the cloud. Even if these inefficiency
factors are important or the mass that can reach the disk outer edge is
smaller, the probability to see the cloud can still be reasonable. For
example, if the cloud were generated for only 100 orbits, requiring a
total mass of just $\sim 10^{-3}\Msun$ to be placed on the
photoevaporating gas streams over all the orbits, then the rate obtained
in equation (13) implies a probability of 1\% to see this cloud at any
random time. This is still a reasonable probability, taking into account
that the probability is calculated a posteriori, after having observed a
curious phenomenon in the Galactic center that might be one among many
unlikely phenomena that could be observed but are not actually happening
at our present time.

  We have noted also in Section 2.4 that collisions between main-sequence
stars might produce similar star-disk systems as tidal encounters with
black holes. The main uncertainty there is that any collision occurring
as close to the Galactic center as the observed cloud would take place
at a very high relative velocity, and it is therefore questionable that
much of the debris that are generated may remain on bound orbits around
one of the two stars. But if a disk of substantial mass can also be
formed in this case, the rate of disk-forming events may be further
increased.

  The observed cloud actually has a complex head-tail structure. The
head is the region of highest surface brightness in the recombination
lines, with a long axis of 100 AU in 2011, which emits the observed
$L$-band infrared emission that appears unresolved. This head is the
cloud we have been
considering in this paper, but there is also a lower surface brightness
tail that is falling behind. This has been interpreted as a large shell
of material from colliding winds that was produced near the orbital
apobothron \citep{S12}, but it may just as well be material that was
barely unbound from the star and was detached from the main cloud near
apobothron, and is now falling behind because of ram-pressure effects.

  The model proposed here can be tested after the cloud passes the
peribothron. The present cloud should be totally disrupted whether
or not a star is contained inside it. Detailed hydrodynamic simulations
predicting the evolution of the disrupted cloud have been presented by
\cite{S12}, which will be very interesting to test the interaction of
the tidal debris from the cloud with the hot medium. However, the test
that will distinguish the model presented here of a star-disk system
inside the cloud is whether a point source emitting in the infrared
continuum and recombination lines remains on the unaltered Keplerian
orbit after the peribothron passage.

  In fact, long after the peribothron passage, a cloud similar to the
present one should be regenerated around the star. The new cloud is
likely to be initially small, and therefore faint in recombination line
emission (which is proportional to the cloud area if it arises from the
external photoionizing radiation). It might therefore be difficult to
distinguish from the surrounding complex debris of the tidally disrupted
cloud, but eventually it should appear as a region of higher surface
brightness in the recombination lines on the exactly predicted position.
Precisely what may happen is difficult to predict and depends on the
structure of the disk and the mass of gas streams that are launched from
it. Hydrodynamic simulations of the process are required for any
quantitative predictions. However, a reasonable expectation is that the
tidally induced internal shocks in the disk during the peribothron
passage may produce enough heating to cause a substantial brightening of
the infrared source. For example, if a mass of $10^{-3}\Msun$ is present
near the outer edge of the disk and is shocked at velocities of $\sim 30
\kms$, the energy released can be up to $10^{43}$ ergs and the disk may
radiate at $\sim 100 \Lsun$ during several months after the peribothron
passage with a surface temperature near 2000 K, implying a very large
brightening in the $K$-band. After the peribothron passage, the observed
light curve of the source in both recombination lines and infrared
continuum should provide a detailed diagnostic of the process of tidal
perturbation and mass ejection from the disk.

  Finally, an interesting consequence of the model presented here is the
possibility that planets are formed in the disks generated in these
encounters. This possibility is basically a revival of the previous
encounter theory for the formation of the solar system, in which planets
were formed from the tidal plume generated during an encounter between
two stars. In environments of a very high stellar density such as the
Galactic center, tidal encounters occur at a high rate to form planetary
systems around a large fraction of stars over the age of the Galaxy.
The problems of the encounter theory \citep{R35,S39} can be overcome:
planets need to form only after the ejected material has formed a disk
around the star and cooled down, and the fast rotation of the star
can expand the disk and provide angular momentum. It is interesting
that planets may have formed in the hypothesized disk within the
observed cloud in the Galactic center, and that a type of planetary
systems of this special origin might be present in the stellar cusp
surrounding $\sgra$.

\acknowledgments

  I thank Charles Gammie and Andy Gould for helpful discussions, and
Scott Tremaine for pointing out the relation of this model to the
encounter theory for planet formation. I am grateful to the Department
of Astronomy at Pennsylvania State University for their hospitality
during the time this work was carried out. This work has been supported
in part by the Spanish grants AYA2009-09745 and PR2011-0431.

{}

\end{document}